\documentclass[notitlepage]{revtex4-1}
\usepackage[legalpaper, margin=1.15in]{geometry}
\usepackage{setspace}
\usepackage[utf8]{inputenc}
\usepackage[colorlinks=true,citecolor=blue,linkcolor=blue]{hyperref}
\usepackage[normalem]{ulem}
\usepackage{url}
\usepackage{graphicx,wrapfig,float,slashed,cancel}
\usepackage{amsmath,amssymb,epsfig,xcolor,stmaryrd}
\usepackage{bm}
\usepackage{enumitem}
\usepackage{hhline,multirow,tabularx}  
\usepackage{graphics}
\usepackage{latexsym}
\usepackage{rotating}
\usepackage{subfigure}
\usepackage{color}
\usepackage{blindtext}
\usepackage{lipsum}
\usepackage{physics}

\begin{document}


\title{Impact of JLab data on the determination of GPDs at zero skewness and new insights from transition form factors $ N\rightarrow \Delta $} 

\author{The MMGPDs\footnote{Modern Multipurpose GPDs} Collaboration:\\
Muhammad Goharipour$^{1,2}$}
\email{muhammad.goharipour@ipm.ir}
\thanks{Corresponding author}

\author{Hadi Hashamipour$^{2}$}
\email{h\_hashamipour@ipm.ir}

\author{Fatemeh Irani$^{1}$}
\email{f.irani@ut.ac.ir }

\author{K.~Azizi$^{1,2,3}$}
\email{kazem.azizi@ut.ac.ir}


\affiliation{
$^{1}$Department of Physics, University of Tehran, North Karegar Avenue, Tehran 14395-547, Iran\\
$^{2}$School of Particles and Accelerators, Institute for Research in Fundamental Sciences (IPM), P.O. Box 19395-5531, Tehran, Iran\\
$^{3}$Department of Physics, Do\v{g}u\c{s} University, Dudullu-\"{U}mraniye, 34775
Istanbul, Turkey
}

\date{\today}

\preprint{}

\begin{abstract}

It is well established now that the generalized parton distributions (GPDs) at zero skewness are playing important roles in some physical process such as elastic electron-nucleon scattering, elastic (anti)neutrino-nucleon scattering, and wide-angle Compton scattering (WACS) via various types of form factors (FFs). In this study, we are going to utilize the recent JLab measurements of the elastic electron-nucleon scattering reduced cross-section, namely GMp12, as a touchstone to unravel the tension observed between the measurements of the WACS cross-section and the data of the proton magnetic FF $ G_M^p $. We also investigate the impact of GMp12 data on valence unpolarized GPDs $ H_v^q $ and $ E_v^q $ at zero skewness by performing some $ \chi^2 $ analyses of the related experimental data. By calculating the electric and scalar quadrupole ratios, $ R_{EM} $ and $ R_{SM} $, and magnetic transition FF $ G^*_M/3G_D $ related to the nucleon-to-delta ($ N\rightarrow \Delta $) transition using the extracted GPDs and comparing the results with corresponding experimental measurements, we show that our results are in an excellent consistency with experiments, indicating the universality property of GPDs. We emphasize that the inclusion of these data in the future analysis of GPDs can significantly affect the extracted GPDs especially their uncertainties at smaller values of $ Q^2 $.

\end{abstract}


\maketitle

\renewcommand{\thefootnote}{\#\arabic{footnote}}
\setcounter{footnote}{0}

\section{Introduction}\label{sec:one} 

Parton distribution functions (PDFs) serve as essential inputs for theoretical calculations of physical observables in high-energy processes involving hadrons such as deep inelastic scattering (DIS) and hadron-hadron collisions which provide the precision test of the Standard Model (SM), understanding the QCD dynamics, and searches for new physics~\cite{Forte:2013wc,Ethier:2020way,Rojo:2019uip}. However, PDFs just describe the longitudinal momentum distribution of partons within hadrons and hence can not provide any information in the transverse plane. In this context, one can refer to the transverse momentum distributions (TMDs)~\cite{Close:1977mr,Kimber:2001sc,Angeles-Martinez:2015sea,Signal:2021aum,Bacchetta:2022awv,Boussarie:2023izj} and GPDs~\cite{Muller:1994ses,Radyushkin:1996nd,Ji:1996nm,Ji:1996ek,Burkardt:2000za} to get further information on the 3D structure of hadrons~\cite{Diehl:2015uka}. The first ones give correlations between the transverse and longitudinal momentum of partons inside a hadron, and the latter ones represent correlations between the transverse position and the longitudinal momentum of partons. The hadron structure can be described even in more details through the Wigner distribution~\cite{Wigner:1932eb} which encodes all the nonperturbative QCD dynamics of parton constituents and hence represents a comprehensive visualization of the partonic structure of hadron in five dimensions~\cite{Belitsky:2003nz,Lorce:2011ni,Mukherjee:2015aja,Liu:2015eqa,Chakrabarti:2017teq,Pasechnik:2023mdd}.

Among different objects that describe the nonperturbative QCD dynamics of partons inside the nucleon ($ p $ and $ n $ denote the proton and neutron, respectively, throughout this article), GPDs are of special importance because of their connections with other objects~\cite{Diehl:2003ny,Boffi:2007yc,Diehl:2015uka}. On one hand, unpolarized (polarized) PDFs $ q(x,\mu^2) $ ($ \Delta q $), where $ x $ is the longitudinal momentum fraction of the nucleon carried by partons and $ \mu $ is the factorization scale at which the partons are resolved, can be obtained from unpolarized (polarized) GPDs $ \mathcal{F}^q(x,\mu^2,\xi,t) $ ($ \mathcal{\tilde{F}}^q $) under the so-called forward limit $ t=0 $ and $ \xi=0 $. Here, $ \xi $ is the longitudinal momentum transfer, known as skewness, and $ t=-Q^2 $ is the negative momentum transfer squared. At zero skewness, a Fourier transform of GPDs with respect to the transverse part of the momentum transfer yields the impact parameter distribution $ f(x,\bm{b}) $, which gives the probability density of finding a parton with longitudinal momentum fraction $ x $ at a transverse distance $ \bm{b} $ from the center of momentum of the proton~\cite{Burkardt:2000za,Burkardt:2002hr}. Note that integrating the Wigner distribution $ W(x,\bm{k},\bm{b}) $ over the transverse momentum $ \bm{k} $ yields also $ f(x,\bm{b}) $, that reveals the connection between GPD and Wigner distribution in turn, and integrating $ f(x,\bm{b}) $ over $ \bm{b} $ leads to the corresponding PDF~\cite{Diehl:2015uka}. On the other hand, different Mellin moments of GPDs are associated with different hadron FFs~\cite{Muller:1994ses,Radyushkin:1996nd,Ji:1996nm,Ji:1996ek,Burkardt:2000za,Bernard:2001rs,Pagels:1966zza,Guidal:2004nd} including the electromagnetic FFs (EMFFs), axial FFs (AFFs), gravitational FFs (GFFs), and transition FFs (TFFs). These properties make GPDs an essential ingredient of different types of hard exclusive and elastic scattering processes (to get more information see Refs.~\cite{Goeke:2001tz,Belitsky:2005qn,Kumericki:2016ehc,Hashamipour:2022noy} and references therein).

In recent years, there have been many efforts to study different kinds of GPDs and determine them both theoretically and phenomenologically~\cite{Kumericki:2016ehc,Hashamipour:2022noy,Goloskokov:2007nt,Kumericki:2009uq,Gonzalez-Hernandez:2012xap,Guidal:2013rya,Diehl:2013xca,Selyugin:2014sca,Berthou:2015oaw,Kriesten:2019jep,Hashamipour:2019pgy,Hashamipour:2020kip,Hashamipour:2021kes,Irani:2023lol,Kriesten:2021sqc,Xu:2021wwj,Ahmady:2021qed,CLAS:2022iqy,Guo:2022upw,Guo:2023ahv,Qiu:2022pla,Fu:2022bpf,Duplancic:2023kwe,Constantinou:2020hdm,Riberdy:2023awf,Bhattacharya:2023ays,Bhattacharya:2023nmv,Bhattacharya:2023jsc,Selyugin:2023hqu,Luan:2023lmt,Kaur:2023lun,Kaur:2023zhn,Cichy:2023dgk,Moffat:2023svr,Duplancic:2022ffo,Shuryak:2023siq,Liu:2024umn,Pedrak:2017cpp,Pedrak:2020mfm,Grocholski:2022rqj}. However, there are
still many questions that we should answer and new ideas we can advance. Focusing on the case of zero-skewness GPDs, in Ref.~\cite{Hashamipour:2022noy}, the authors were determined three kinds of GPDs, namely $ H^q(x,t) $, $ E^q(x,t) $, and $ \tilde{H}^q(x,t) $, by performing a simultaneous analysis of a wide range of the elastic scattering experimental data. They showed that there is a significant tension between the data of the wide-angle Compton scattering (WACS) cross section~\cite{Danagoulian:2007gs} and the proton magnetic FF ($ G_M^p $) at
larger values of $ -t $, so that it is not possible to find a set of GPDs that provides a desirable description of them simultaneously. Consequently, they suggested either reassessing the experimental measurements of both $ G_M^p $ and WACS or revising the theoretical calculations of the WACS cross section. In Ref.~\cite{Irani:2023lol}, by considering the new measurement of the antineutrino-proton scattering cross section from the MINERvA Collaboration~\cite{MINERvA:2023avz}, the authors showed that such a tension (but milder) is also observed between the MINERvA and WACS data. They concluded that the MINERvA data do not recommend the exclusion of the $ G_M^p $ data from the analysis, while the WACS data do this.

In the present study, considering the new precision measurements of the elastic electron-proton scattering cross section at high-energy performed in Hall A of Jefferson Lab~\cite{Christy:2021snt}, referred to as GMp12, we are trying to unravel the issue raised in~\cite{Hashamipour:2022noy} concerning the tension between the WACS and $ G_M^p $ data in Sec.~\ref{sec:two}. Then, we investigate the impact of GMp12 data on the unpolarized valence GPDs $ H_v^q(x,t) $ and $ E_v^q(x,t) $ and their uncertainties by performing a standard $ \chi^2 $ analysis of all available elastic electron-nucleon scattering data in Sec.~\ref{sec:three}. The extracted GPDs are used to calculate the electric and scalar quadrupole ratios, $ R_{EM} $ and $ R_{SM} $, and magnetic transition FF $ G^*_M/3G_D $ related to the nucleon-to-delta transition in Sec.~\ref{sec:four}. The results are also compared to the available experimental data and some theoretical predictions. Finally, we summarize our results and conclusions in Sec.~\ref{sec:five}.

\section{WACS VS $ G_M^p $ data}\label{sec:two}

It is well established now that the study of the elastic electron-nucleon scattering can provide crucial information about the electromagnetic nucleon FFs and hence the distribution of charge and magnetic moments within the nucleon~\cite{Rosenbluth:1950yq,Yennie:1957skg,Hand:1963zz,Chen:2023dxp}. The precise measurements of such scattering processes are essential for testing theoretical models, particularly the electroweak sector of the SM, and improving our understanding of the underlying physics.
Elastic electron-nucleon scattering experiments contribute also to the determination of the root-mean-square  radii of the proton and neutron. From past to present, there have been many measurements of this process from different experiments as well as many efforts to extract the electromagnetic FFs or their ratios from the original cross sections~\cite{Christy:2021snt,Kelly:2002if,JeffersonLaboratoryE93-038:2005ryd,Arrington:2007ux,Qattan:2012zf,A1:2013fsc,Ye:2017gyb,McRae:2023zgu}. 

Recently, GMp12 experiment which performed at Hall A of Jefferson Lab using the basic suite
of experimental instrumentation~\cite{Alcorn:2004sb}, has presented new precision measurements of the elastic electron-proton scattering cross section for momentum transfer squared ($ -t=Q^2 $) up to 15.75 GeV$ ^2 $~\cite{Christy:2021snt}. Considering the one-photon exchange approximation, the differential elastic electron-nucleon scattering cross section can be written as the product of Mott cross section and a structure-dependent term that depends on the electric and magnetic FFs (called Sachs FFs), $ G_E $ and $ G_M $, as follows~\cite{Sachs:1962zzc},
\begin{align}
 \frac{d\sigma}{d \Omega} = \left(
\frac{d\sigma}{d\Omega} \right)_{\rm Mott} \frac{\epsilon G_E^2(Q^2) + \tau
	G_M^2(Q^2)}{\epsilon (1+\tau)} \,.
\label{Eq1}
\end{align}
In above equation, Sachs FFs are functions of $ Q^2=4EE^\prime \sin^2(\theta/2) $ where $ \theta $ is the scattering angle of the electron while $ E $ and $ E^\prime $ are its incident and scattered energies in lab frame. Moreover, the dimensionless kinematic variables $\epsilon$ and $\tau$ are written as
\begin{equation}
\label{Eq2}
\begin{split}
	\tau = \frac{Q^2}{4m^2} \,, \quad
	\epsilon = \left[ 1 + 2 (1 + \tau) \tan^2 \frac{\theta}{2} \right]^{-1} \,,
\end{split}
\end{equation}
where $ m $ is the nucleon mass. Note that the fine-structure constant $ \alpha $ appears in the Mott cross section which is the cross section for a recoil-corrected relativistic structureless object,
\begin{align}
\label{Eq3}
\left( \frac{d\sigma}{d\Omega} \right)_{\rm Mott}
=\frac{\alpha^2}{4 E^2 \sin^4{\theta / 2}} \frac{E^\prime}{E}\cos^2{\frac{\theta}{2}} \,. 
\end{align}
Note that $G_E^p(0) = 1$, $G_E^n(0) = 0$, and $G^j_M(0) = \mu_j $, where $ \mu_j $ is the magnetic moment of the nucleon $ j=p,n $. It is also common to express the structure-dependent term in Eq.~(\ref{Eq1}), isolated in the reduced cross
section $ \sigma_R $, in terms of the longitudinal and transverse contributions to the cross section, $ \sigma_T $ and $ \sigma_L $~\cite{Christy:2021snt},
\begin{eqnarray}
\label{Eq4}
\nonumber
\sigma_R &=& \epsilon G_E^2(Q^2) + \tau G_M^2(Q^2)=\sigma_T + \epsilon \sigma_L\,, \\
 &=& G_M^2(Q^2)[\tau +\epsilon {\rm RS}(Q^2)/\mu_j^2]\,,
\end{eqnarray}
where $ {\rm RS}(Q^2)=(\mu_jG_E/G_M)^2 $ is the normalized Rosenbluth slope.

The Sachs FFs in turn are expressed in terms of the Dirac and Pauli FFs of the nucleon, $ F_1 $ and $ F_2 $, 
\begin{align}
\label{Eq5}
G_M(Q^2) &= F_1(Q^2) + F_2(Q^2) \,, \nonumber \\ 
G_E(Q^2) &= F_1(Q^2) - \frac{Q^2}{4m^2} F_2(Q^2) \,.
\end{align}
On the other hand, considering the proton case, the Dirac and Pauli FFs are related to the unpolarized valence GPDs $ H_v^q $ and $ E_v^q $ at zero skewness ($ \xi=0 $), where $ q=u,d $ stands for the up and down quarks (the strange-quark contribution can be neglected), using the following sum rules~\cite{Diehl:2004cx,Diehl:2013xca}
\begin{align}
F^p_1(Q^2)=\sum_q e_q F^q_1(Q^2)=\sum_q e_q \int_{0}^1 dx\, H_v^q(x,\mu^2,Q^2)\,, \nonumber \\ 
F^p_2(Q^2)=\sum_q e_q F^q_2(Q^2)=\sum_q e_q \int_{0}^1 dx\, E_v^q(x,\mu^2,Q^2)\,,
\label{Eq6}
\end{align}
where $ e_q $ refers to the electric charge of the constituent quark $ q $. On can obtain the related expressions for the neutron FFs $ F_1^n $ and $ F_2^n $ considering the isospin symmetry $ u^p=d^n, d^p=u^n $.

According to the above equations, by measuring the elastic electron-nucleon scattering cross section or equivalently the electromagnetic FFs, one can get important information about the unpolarized valence GPDs $ H_v^q $ and $ E_v^q $ at zero skewness. Actually, such measurements play the main role in the extraction of the unpolarized GPDs and the measurements of the WACS cross section or the elastic (anti)neutrino-proton scattering cross-section play a complementary role~\cite{Hashamipour:2022noy,Irani:2023lol}. So, it is of interest to investigate the impact of new GMp12 measurements~\cite{Christy:2021snt} on $ H_v^q $ and $ E_v^q $ GPDs. We will do this mission in the next section. The main question we are going to focus on in this section is how do GMp12 measurements judge about the tension observed between the WACS cross-section measurements~\cite{Danagoulian:2007gs} and $ G_M^p $ data explained before. With a lesser importance, it is also of interest to address the tension observed in Ref.~\cite{Hashamipour:2022noy} between the world data of $ G_M^p $ from AMT07~\cite{Arrington:2007ux} analysis and the corresponding ones from Mainz~\cite{A1:2013fsc} data.
To answer these questions, we make some predictions of the GMp12 data using different GPD sets determined in Ref.~\cite{Hashamipour:2022noy} in order to investigate how the theoretical predictions match the GMp12 data. It is worth noting in this context that the GMp12 data have really a merit to unravel these issues because the measurements have been presented not only for $ G_M^p $ but also for the reduced cross-section $ \sigma_R $ in Eq.~(\ref{Eq4}) which includes both $ G_E $ and $ G_M $. In fact, if there is something wrong with the $ G_M^p $ data versus the WACS cross-section measurements, the GPD sets obtained by including the $ G_M^p $ data in the analysis would not be able to describe the $ \sigma_R $ GMp12 data well.

Now, let us briefly introduce different GPD sets from Ref.~\cite{Hashamipour:2022noy} that we are going to use them to make theoretical predictions of the GMp12 measurements of $ \sigma_R $. 
\begin{itemize}
 \item Set 2: this set has been obtained by analyzing data of the ratio of the proton electric and magnetic FFs $, R^p=\mu_p G_E^p/G_M^p $, and the neutron electric and magnetic FFs, $ G_E^n $ and $ G_M^n/\mu_n G_D $, from YAHL18 analysis~\cite{Ye:2017gyb}, the data of the proton magnetic FF, $ G_M^p $, from AMT07 analysis~\cite{Arrington:2007ux}, and finally the data of the charge and magnetic radii of the nucleons~\cite{ParticleDataGroup:2018ovx}.
 \item Set 3: this set has been obtained by analyzing the same data of Set 2 with the difference that AMT07 data of $ G_M^p $ have been replaced with the corresponding Mainz~\cite{A1:2013fsc} data. 
 \item Set 11: this set has been obtained by analyzing all data that were used for Set 2 and Set 3 (both AMT07 and Mainz data have been included) in addition to data of the proton AFF, $ F_A $, from various experiments as well as the CLAS Collaboration measurements at large $ -t $~\cite{CLAS:2012ich}, and finally the WACS cross-section data from the JLab Hall A Collaboration~\cite{Danagoulian:2007gs}. 
 \item Set 12: this set has been obtained by removing the AMT07 and Mainz data of $ G_M^p $ from the analysis of Set 11.
 \end{itemize} 
 
\begin{figure}[!htb]
    \centering
\includegraphics[scale=0.8]{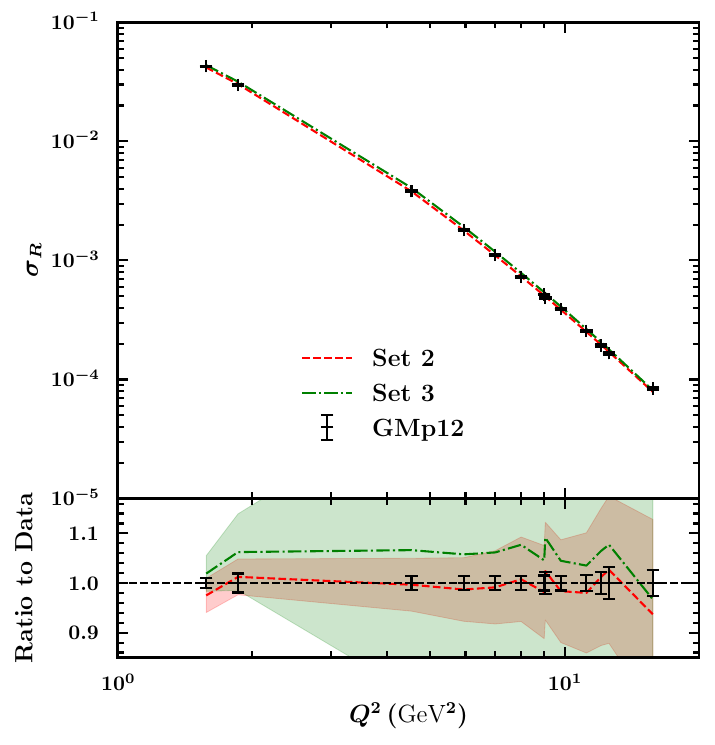}    
    \caption{A comparison between the experimental data from the GMp12 measurements~\cite{Christy:2021snt} for the reduced cross-section of Eq.~(\ref{Eq4}) and the corresponding theoretical predictions obtained   using Set 2 and Set 3 of GPDs taken from Ref.~\cite{Hashamipour:2022noy}.}
\label{fig:IGHA2023-1}
\end{figure}

Figure~\ref{fig:IGHA2023-1} shows a comparison between the GMp12 measurements~\cite{Christy:2021snt} of the reduced cross-section $ \sigma_R $ and the corresponding theoretical calculation obtained using the GPD Set 2 and Set 3 from Ref.~\cite{Hashamipour:2022noy}. Additionally, the ratios of these predictions to the data with their uncertainties have been plotted in the lower panel to investigate the differences more precisely as the value of $ Q^2 $ increases. The results clearly show that Set 2 has the better description of data compared with Set 3. Note that Set 2 has been obtained by removing the $ G_M^p $ Mainz data (which covers only small $ Q^2 $ values) from the analysis and considering only the AMT07 world data (which cover a wide range of the $ Q^2 $ values). This indicates that the new and precise GMp12 measurements are more consistent with the AMT07 data. Considering the tension between the AMT07 and Mainz data at small values of $ Q^2 $, we can conclude that the GMp12 data give more validity to GPDs of Set 2 than Set 3. As a result, one may decide to ignore the Mainz data in the future global analysis of GPDs. Considering the uncertainty bands in the bottom panel of Fig.~\ref{fig:IGHA2023-1}, we can infer that the inclusion of the GMp12 data in the global analysis of GPDs can decrease the GPD uncertainties especially at larger values of $ Q^2 $ as we will show in the next section. Note that the large uncertainties of GPDs Set 3 is a reflection of the fact that there were no any $ G_M^p $ data points at large $ Q^2 $ due to removing the AMT07 world data from the analysis. 

\begin{figure}[!htb]
    \centering
\includegraphics[scale=0.8]{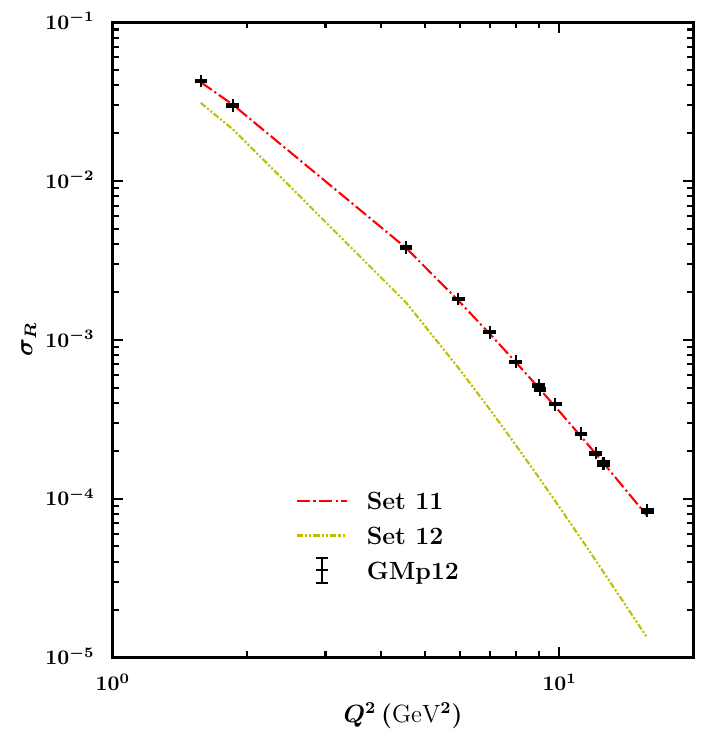}    
    \caption{Same as Fig.~\ref{fig:IGHA2023-1}, but for GPD Set 11 and Set 12.}
\label{fig:IGHA2023-2}
\end{figure}

Figure~\ref{fig:IGHA2023-2} shows the same comparison as Fig.~\ref{fig:IGHA2023-1} but for the results calculated using Set 11 and Set 12 of Ref.~\cite{Hashamipour:2022noy}. According to the results obtained, Set 12 that has been obtained by removing all $ G_M^p $ data from the analysis (both AMT07 and Mainz) has a inadequate description of GMp12 measurements while Set 11 is completely in agreement of the data. Note that the analysis of Set 11 included the WACS and $ G_M^p $ data simultaneously (besides other experimental data) regardless of the tension between them. Although Set 12 has a remarkably better description of the WACS data compared with Set 11 (see Fig. 28 of Ref.~\cite{Hashamipour:2022noy}) it absolutely cannot describe the GMp12 measurements of the reduced cross-section $ \sigma_R $. Considering the fact that here the comparisons have been made with the GMp12 $ \sigma_R $ data instead of its extracted $ G_M^p $ data, we can conclude that there is not something wrong with the previous measurements of $ G_M^p $. Actually, GMp12 data do not recommend the exclusion of the $ G_M^p $ data from the analysis unlike the measurements of the WACS cross-section by JLab~\cite{Danagoulian:2007gs}. Therefore, the tension observed between the WACS and $ G_M^p $ has to have another origin. For instance, it may suggest revising the theoretical calculations of the WACS cross section or improving the phenomenological framework, e.g., by considering the strange quark or gluon contributions or searching the optimum value of the factorization scale $ \mu $ as done in  Refs.~\cite{Hashamipour:2019pgy,Hashamipour:2020kip}.
%

%
%
\section{Impact of GMp12 data on GPDs}\label{sec:three}

According to the findings presented in the previous section, it is of interest now to investigate the impact of new GMp12 data~\cite{Christy:2021snt} on the determination of GPDs. Actually, these data can benefit us in two ways. On one hand, because of the high accuracy of these data, they can lead to a remarkable decrease in the uncertainties of the extracted GPDs though their central values might also be tuned. One the other hand, since these data contain also data points with large $ Q^2 $ value, they can really improve our knowledge in the kinematic where the older world experimental data of $ G_M^p $ have not a considerable contribution and have large uncertainties unlike the small $ Q^2 $ region.  In this section, we first introduce briefly the phenomenological framework we adopt for performing our QCD analysis of GPDs. Then we study the impact of GMp12 data on the unpolarized valence GPDs $ H_v^q $ and $ E_v^q $ by performing two separate analyses, one with and the other without considering the GMp12 data in the analysis, and comparing their results with each other. 

\subsection{Phenomenological framework}

As mentioned in Sec.~\ref{sec:two}, the unpolarized valence GPDs $ H_v^q $ and $ E_v^q $ at zero skewness ($ \xi=0 $) are related to the Sachs FFs, $ G_E $ and $ G_M $, through Eqs.~(\ref{Eq5}) and~(\ref{Eq6}). So, in order to calculate electromagnetic FFs and hence the reduced cross-section $ \sigma_R $ in Eq.~(\ref{Eq4}), one needs a suitable parameterization of GPDs in terms of $ x $ and $ Q^2 $ (or equivalently $ t $). In this study, we adopt the ansatz proposed in Refs.~\cite{Diehl:2004cx,Diehl:2013xca} that, as it has been shown in Refs.~\cite{Hashamipour:2019pgy,Hashamipour:2020kip,Hashamipour:2021kes,Irani:2023lol,Hashamipour:2022noy}, leads to a good description of data, 
\begin{align}
H_v^q(x,\mu^2,t)= q_v(x,\mu^2)\exp [tf_v^q(x)]\,,  \nonumber \\ 
E_v^q(x,\mu^2,t)= e_v^q(x,\mu^2)\exp [tg_v^q(x)]\,.
\label{Eq7}
\end{align}
Here $ q_v(x,\mu^2) $ is the unpolarized PDF for the valence quark $ q $ which is taken from the the next-to-leading order (NLO) \texttt{NNPDF} analysis~\cite{NNPDF:2021njg} at scale $ \mu=2 $ GeV, using the \texttt{LHAPDF} package~\cite{Buckley:2014ana} (as mentioned before, the contribution of the strange quark is neglected in the present study). For the profile functions $ f_q(x) $ and $ g_q(x) $ we use the following parametrization form as suggested in~\cite{Diehl:2004cx,Diehl:2013xca}
\begin{equation}
\label{Eq8}
{\cal F}(x)=\alpha^{\prime}(1-x)^3\log\frac{1}{x}+B(1-x)^3 + Ax(1-x)^2.
\end{equation}
Although the profile function can take other forms, the above form is flexible enough and leads to a better fit of the data~\cite{Hashamipour:2019pgy}. Note also that at forward limit ($ t=0 $ and $ \xi=0 $), GPDs in Eq.~(\ref{Eq7}) reduce to the corresponding PDFs. However, in the case of GPDs $ E_v^q $, there is no parameterization available for the forward limit $ e_v^q(x,\mu^2) $ from a previous determination. Therefore, $ e_v^q(x,\mu^2) $ must also be obtained from the fit to data. In this regard, we consider a parametrization form similar to Ref.~\cite{Diehl:2013xca} that has also been used in the analyses of Ref.~\cite{Hashamipour:2022noy,Hashamipour:2021kes}
\begin{equation}
\label{Eq9}
e_v^q(x,\mu^2)=\kappa_q N_q x^{-\alpha_q} (1-x)^{\beta_q} (1+\gamma_q\sqrt{x})\,,
\end{equation}
which is defined at $ \mu=2 $ GeV. $ \kappa_u $ and $ \kappa_d $ are the magnetic moments of the up and down quarks that can be computed from the measured magnetic moments of proton and neutron in the units of nuclear magneton. In the present study, we use $ \kappa_u=1.67 $ and  $ \kappa_d=-2.03 $ which have been calculated using the values of the magnetic moments of proton and neutron taken from~\cite{ParticleDataGroup:2018ovx}. Then, the normalization factor $ N_q $ can be obtained from the following sum rule:
\begin{equation}
\label{Eq10}
\int_0^1 dx e_v^q(x)=\kappa_q.
\end{equation}

According to the above explanations, the unknown free parameters associated with each quark flavor in this study are $ \alpha^{\prime} $, $ B $, $ A $, $ \alpha $, $ \beta_q $, and $ \gamma $. The optimum values of these parameters are obtained as usual by performing a standard $ \chi^2 $ analysis of the experimental data in which the following function is minimized:
\begin{equation}
\label{Eq11}
\chi^2= \sum_i^n \Big (\frac{{\cal E}_i-{\cal T}_i}{\delta{\cal E}_i} \Big )^2\,. 
\end{equation}
Here $ {\cal E}_i $ is the measured value of the experimental data point $ i $ and $ {\cal T}_i $ denotes the corresponding theoretical prediction. The summation is performed over all data point included in the analysis taken from different datasets. One can obtain the experimental errors $ \delta{\cal E}_i $ associated with each measured data point by adding the systematic and statistical errors in quadrature.
To perform the minimization procedure in this study we use the CERN program library \texttt{MINUIT}~\cite{James:1975dr}. Our method for calculating uncertainties of GPDs and physical observables is as in Ref.~\cite{Pumplin:2001ct} that is the standard Hessian approach. Note that we use the standard value $ \Delta \chi^2=1 $ for constructing the confidence region.

As discussed in Ref.~\cite{Diehl:2013xca}, the the profile functions and forward limits of GPDs must obey a positivity condition which comes from the fact that the quark and
antiquark distributions in $ x $ cannot be same at a nominal transverse distance $ b $ from the proton center. In this study, we adopt the method proposed in~\cite{Hashamipour:2021kes} (Scenario 2) in which the positivity condition is preserved automatically in a wide range of the $ x $ values by incorporating condition $ g_v^q(x) < f_v^q(x) $ in the main body of the fit program.
It is well established now that a suitable and efficient approach to determine the optimum values of the unknown parameters in every global analysis of the experimental data is utilizing the parametrization scan procedure~\cite{Hashamipour:2021kes,Hashamipour:2022noy,Irani:2023lol} proposed first in the QCD analysis of PDFs performed by H1 and ZEUS Collaborations~\cite{H1:2009pze}. An advantage of this procedure is that the final parametrization of each distribution is obtained systematically. Accordingly, here we follow the same approach to find the optimum parameterizations and the best values of their unknown parameters.

\subsection{Data selection}\label{sec:three-two}

Just as for the analyses of Refs.~\cite{Hashamipour:2021kes,Hashamipour:2022noy}, the main body of the experimental data that we consider in the present study comprises of the YAHL18 data~\cite{Ye:2017gyb} which include a wide and updated range of the world electron scattering data off both proton and neutron targets and have been obtained by incorporating the two-photon exchange (TPE) corrections.
These data include 69, 38, and 33 data points of the world  $ R^p=\mu_p G_E^p/G_M^p $ polarization, $ G_E^n $, and $ G_M^n/\mu_n G_D $ measurements, respectively. Here, $ G_D=(1+Q^2/\Lambda^2)^{-2} $ is the dipole form factor with $ \Lambda^2=0.71 $ GeV$ ^2 $. Overall, the YAHL18 data cover the  $ -t $ range from 0.00973 to 10 GeV$ ^{2} $ for 140 data points. Since GPDs $ H_v^q $ and $ E_v^q $ at zero skewness are related to the mean squared of the charge and magnetic radii of the nucleons, $ r_{jE} $ and $ r_{jM} $, through the Sachs FFs 
\begin{align}
\left<r_{jE}^2\right>= \left.  6 \dv{G_E^j}{t} \right|_{t=0} \,, \nonumber \\ 
\left<r_{jM}^2\right>= \left.  \frac{6}{\mu_j} \dv{G_M^j}{t} \right|_{t=0},
\label{Eq12}
\end{align}
we also consider their related measurements (four data points) from~\cite{ParticleDataGroup:2018ovx}. Note that these data can provide crucial information about the small-$ t $ behavior of GPDs, especially the profile functions $ f_q(x) $ and $ g_q(x) $. In addition to these data, in order to put further constraints on GPDs, we include the data of $ G_E^p $ from AMT07~\cite{Arrington:2007ux} and Mainz~\cite{A1:2013fsc} which contain 47 and 77 data points, respectively. These data cover the  $ -t $ range from 0.007 to 5.85 GeV$ ^{2} $. An important point in this context that should be mentioned is that we could use instead the data of $ G_M^p $ from these analyses, but as it has been shown in Refs.~\cite{Hashamipour:2022noy,Ye:2017gyb} the tension between the world AMT07 data and Mainz measurements is worse in this case. Hence, we decided to use the $ G_E^p $ data in the present study which show less tension between AMT07 and Mainz data~\cite{Ye:2017gyb}. Finally, as mentioned before, we include the new data of the reduced cross-section $ \sigma_R $ from JLab experiment~\cite{Christy:2021snt} for the first time in order to study their impact on the extracted GPDs.
These data cover the  $ -t $ range from 1.577 to 15.76 GeV$ ^{2} $. So, considering also the fact that they are precise compared with the older data, they provide crucial information on GPDs especially at larger values of $ -t $.

\subsection{Results}\label{sec:three-three}

As discussed before, one of the main goals of the present study is investigating the impact of the new and precise measurements of the reduced cross-section of the elastic electron-proton scattering defined in Eq.~(\ref{Eq4}) from JLab experiment~\cite{Christy:2021snt}, namely GMp12, on the unpolarized valence GPDs at zero skewness. In Sec.~\ref{sec:two}, we showed that these data are more consistent with GPDs Set 2 of Ref.~\cite{Hashamipour:2022noy} where the Mainz data have been excluded from the analysis and also GPDs Set 11 where the $ G_M^p $ data (both AMT07 and Mainz) have not been excluded from the analysis and have been analyzed besides the AFFs and WACS data.
Now it is of interest to see how the GMp12 data affect the extracted unpolarized valence GPDs $ H_v^q $ and $ E_v^q $ especially at larger values of $ -t $ where we expect that these data have grater effect.
To this aim, here we construct three analyses of GPDs. The first analysis which is called the ``Base Fit" includes all data sets described in Sec.~\ref{sec:three-two} except GMp12 data. In this way, we obtain a set of GPDs which has no influence from the GMp12 data. Note that it is different from Set 1 of~\cite{Hashamipour:2022noy} (where both AMT07 and Mainz data have been included in the analysis) since we are going to use the $ G_E^p $ data instead of $ G_M^p $ as explained before. Then, we repeat the Base Fit but by including also the GMp12 data in the analysis to investigate their impact on the extracted GPDs. The GPD set that is obtained from this analysis is called ``GMp12". In addition to these analyses, we perform an analysis by removing the Mainz data considering the notable tension observed between the AMT07 and Mainz data~\cite{Ye:2017gyb} (note again that this tension is absolutely significant for the case of $ G_M^p $ data). This set is called ``GMp12\_NoMainz".

Following the parametrization scan procedure as described in~\cite{Hashamipour:2021kes,Hashamipour:2022noy,Irani:2023lol}, we find a set of GPDs with $ \alpha^{\prime}_{g_v^u}=\alpha^{\prime}_{f_v^u} $, $ B_{g_v^u}=0 $, and $ B_{g_v^d}=0 $.
Actually, releasing these parameters does not lead to any decrease in the value of $ \chi^2 $ and hence any improvement in the quality of the fit.
All other parameters are obtained from the fit to data as usual. Table~\ref{tab:par} contains
the optimum values of the parameters obtained from three analyses described above. As can be seen, 
the inclusion of the GMp12 data in the analysis has led to a significant decrease in uncertainties of the parameters as expected. Their central values have also been change significantly in some cases. Note that removing the Mainz data from the analysis does not changed considerably the final results, especially for GPDs $ H_v^q $. However, some changes are seen in the forward limits $ e_v^q(x) $ of GPDs $ E_v^q $. The impact of GMp12 data on GPDs will be investigated later in more details where we compare the extracted GPDs to each other.
\begin{table}[th!]
\scriptsize
\setlength{\tabcolsep}{8pt} 
\renewcommand{\arraystretch}{1.4} 
\caption{The optimum values of the parameters of the profile functions~(\ref{Eq8}), and distributions $ e_v^q(x) $ of Eq.~(\ref{Eq9}) obtained from three analyses described in Sec.~\ref{sec:three-three}.}\label{tab:par}
\begin{tabular}{lcccc}
\hline
\hline
 Distribution &  Parameter           &  Base Fit           &  GMp12            &  GMp12\_NoMainz  \\
\hline 
\hline
$ f_v^u(x) $  & $ \alpha^{\prime} $  & $ 0.721\pm0.012 $ & $ 0.712\pm0.008 $ & $ 0.708\pm0.009 $  \\
			  &	$ A $                & $ 0.987\pm0.123 $ & $ 0.941\pm0.039 $ & $ 0.936\pm0.040 $  \\
			  &	$ B $                & $ 0.830\pm0.057 $ & $ 0.864\pm0.030 $ & $ 0.876\pm0.034 $  \\
\hline
$ f_v^d(x) $  & $ \alpha^{\prime} $  & $ 0.412\pm0.037 $ & $ 0.396\pm0.028 $ & $ 0.392\pm0.027 $  \\
			  &	$ A $                & $ 1.644\pm0.539 $ & $ 1.559\pm0.365 $ & $ 1.557\pm0.348 $  \\
			  &	$ B $                & $ 1.489\pm0.205 $ & $ 1.556\pm0.148 $ & $ 1.567\pm0.141 $  \\
\hline
$ g_v^u(x) $  & $ \alpha^{\prime} $  & $ \alpha^{\prime}_{f_v^u} $ & $ \alpha^{\prime}_{f_v^u} $ & $ \alpha^{\prime}_{f_v^u} $  \\
			  &	$ A $                & $ 0.577\pm0.295 $ & $ 0.381\pm0.154 $ & $ 0.302\pm0.184 $  \\
			  &	$ B $                & $ 0.000         $ & $ 0.000 $         & $ 0.000 $  \\
\hline
$ g_v^d(x) $  & $ \alpha^{\prime} $  & $ 0.735\pm0.058 $ & $ 0.730\pm0.057 $ & $ 0.733\pm0.057 $  \\
			  &	$ A $                & $ 1.773\pm0.512 $ & $ 1.710\pm0.460 $ & $ 1.709\pm0.389 $  \\
			  &	$ B $                & $ 0.000 $         & $ 0.000 $         & $ 0.000 $  \\
\hline
$ e_v^u(x) $  & $ \alpha $           & $ 0.648\pm0.054 $ & $ 0.678\pm0.041 $ & $ 0.677\pm0.055 $  \\
			  &	$ \beta $            & $ 7.114\pm1.111 $ & $ 7.537\pm0.610 $ & $ 7.904\pm0.812 $  \\
			  &	$ \gamma $           & $ 2.232\pm2.617 $ & $ 3.301\pm2.278 $ & $ 3.648\pm3.415 $  \\
\hline
$ e_v^d(x) $  & $ \alpha $           & $ 0.739\pm0.035 $ & $ 0.753\pm0.025 $ & $ 0.748\pm0.035 $  \\
			  &	$ \beta $            & $ 5.778\pm1.038 $ & $ 6.101\pm0.985 $ & $ 5.955\pm0.810 $  \\
			  &	$ \gamma $           & $ 6.649\pm3.992 $ & $ 8.942\pm2.855 $ & $ 7.880\pm4.503 $  \\
\hline 		 	
\hline 	
\end{tabular}
\end{table}

The $ \chi^2 $ values of three analyses described above are presented in Table~\ref{tab:chi2} where we have listed the data sets used in the analysis with their references in the first column. The data have been separated according to their related observables. The range of $ -t $ associated to each data set has been reported in the second column. Note that the GMp12 data cover a wide range of $ -t $ compared with other datasets. This clearly indicates the importance of these data for constraining GPDs.
For each dataset, we have presented the value of $ \chi^2 $ per number of data points, $\chi^2$/$ N_{\textrm{pts.}} $ which makes the deduction on the quality of the fit easier.
The values of total $ \chi^2 $ divided by the number of degrees of freedom, $\chi^2 /\mathrm{d.o.f.} $,
have also been presented in the last row.
\begin{table}[th!]
\scriptsize
\setlength{\tabcolsep}{8pt} 
\renewcommand{\arraystretch}{1.4} 
\caption{The results of three analyses described in Sec.~\ref{sec:three-three}.}\label{tab:chi2}
\begin{tabular}{lcccc}
\hline
\hline
  Observable         &  -$t$ (GeV$^2$)   &  \multicolumn{3}{c}{ $\chi^2$/$ N_{\textrm{pts.}} $  }  \\
                     &                   &     Base Fit    &      GMp12     &     GMp12\_NoMainz    \\
\hline 		 	
\hline 
$G_{E}^p$~\cite{A1:2013fsc}                        & $0.0152-0.5524$& $219.1 / 77$  & $ 220.0 / 77 $ &      $  -  $ \\
$G_E^p/G_D$~\cite{Arrington:2007ux}                & $0.007-5.85$   & $40.1 / 47$   & $ 39.9 / 47$  &       $ 39.9 / 47 $        \\
$R^p = \mu_p G_{E}^p / G_{M}^p$~\cite{Ye:2017gyb}  & $0.162-8.49$   & $102.7 / 69$  & $104.9 / 69$  &  $104.9 / 69$ \\
$G_{E}^n$~\cite{Ye:2017gyb}                        & $0.00973-3.41$ & $27.3 / 38$   & $ 26.8 / 38$  &      $ 26.0 / 38$ \\
$G_M^n/\mu_n G_D$~\cite{Ye:2017gyb}                & $0.071-10.0$   & $38.9 / 33$   & $ 37.4 / 33 $ &  $36.8 / 33 $ \\
GMp12 $\sigma_R$~\cite{Christy:2021snt}            & $1.577-15.76$  & $    -    $   & $15.6 / 13$   &      $ 13.7 / 13$ \\
$\sqrt{\left<r_{pE}^2\right>}$~\cite{ParticleDataGroup:2018ovx}   & $ 0 $    & $1.8 / 1$     & $1.6 / 1 $    &  $0.0 / 1 $     \\
$\sqrt {\left<r_{pM}^2\right>}$~\cite{ParticleDataGroup:2018ovx}  & $ 0 $    & $2.5 / 1$     & $2.4 / 1 $    &  $2.5 / 1 $   \\
$\left<r_{nE}^2\right>$~\cite{ParticleDataGroup:2018ovx}          & $ 0 $    & $0.1 / 1$     & $0.8 / 1 $    &  $0.9 / 1 $   \\
$\sqrt {\left<r_{nM}^2\right>}$~\cite{ParticleDataGroup:2018ovx}  & $ 0 $    & $15.9 / 1$    & $18.0 / 1 $   &  $16.9 / 1 $   \\	
\hline
Total $\chi^2 /\mathrm{d.o.f.} $ &                & $448.4 / 253$ & $467.4 / 266$   &  $241.6 / 189$   \\
\hline
\hline
\end{tabular}
\end{table}

On one hand, considering the results of the Base Fit in Table~\ref{tab:chi2}, a remarkable tension is seen between the $ G_E^p $ data of AMT07 ($\chi^2$/$ N_{\textrm{pts.}}=0.85 $) and Mainz ($\chi^2$/$ N_{\textrm{pts.}}=2.85 $), though it is much more moderate compared with the case of $ G_M^p $ data reported in Ref.~\cite{Hashamipour:2022noy} (2.03 for AMT07 versus 6.02 for Mainz). On the other hand, comparing the results of three analyses in Table~\ref{tab:chi2}, one realizes that the inclusion of the GMp12 data in the analysis does not significantly affect the $ \chi^2 $ values of the other datasets. Therefore, considering the good $ \chi^2 $ values of the GMp12 data in the two last analyses, we can conclude that the GMp12 data give more credit to the AMT07 data than Mainz data (as it was deduced from Fig.~\ref{fig:IGHA2023-1}). Note again that the precise GMp12 data are as the reduced cross-section $ \sigma_R $, i.e., the combination of $ G_E $ and $ G_M $. So, they are a good benchmark to judge about the tension observed between the AMT07 and Mainz data. However, it should be noted that in Ref.~\cite{Hashamipour:2022noy} the authors concluded that the inclusion of the $ G_M^p $ Mainz data in the analysis could be useful because they lead to a better description of the nucleon radii data.

\begin{figure}[!htb]
    \centering
\includegraphics[scale=1.3]{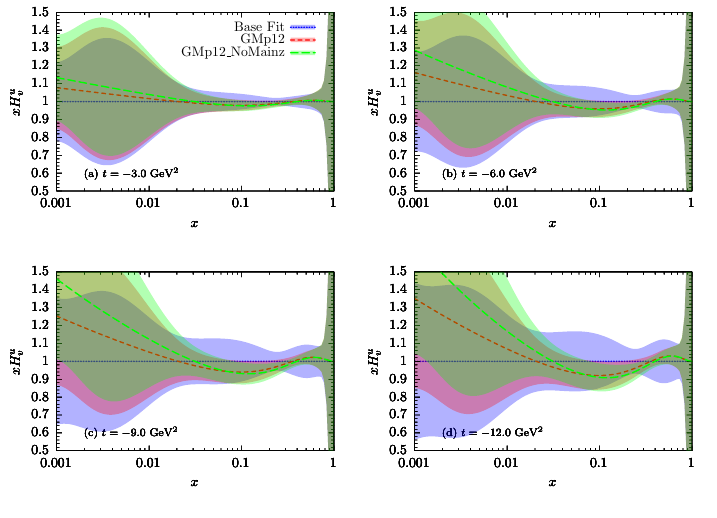}   
    \caption{A comparison between the results of three analyses performed in Sec.~\ref{sec:three-three} for the valence unpolarized GPD $ xH_v^u(x) $ as the ratio to the Base Fit at four $ t $ values shown in panels (a) $ t=-3$, (b) $t=-6$, (c) $t=-9$, and (d) $t=-12 $ GeV$ ^2 $.}
\label{fig:Huv}
\end{figure}

Figure~\ref{fig:Huv} indicates the results obtained for the valence unpolarized GPD $ xH_v^u(x) $ with their uncertainties from three aforementioned analyses. The distributions have been compared as ratio to Base Fit at four different values of $ t $, namely $ t=-3,-6,-9,-12 $ GeV$ ^2 $, in order to study the impact of GMp12 data on $ xH_v^u(x) $ more clearly in a wide range of $ -t $. Note that the error bands include also the uncertainties of the \texttt{NNPDF} PDFs~\cite{NNPDF:2021njg} all distributions at $ t=0 $ are reduced to the $ u_v(x) $ PDF of \texttt{NNPDF}. As expected, the GMp12 data have more impact on $ xH_v^u(x) $ as the absolute value of $ -t $ increases. They lead to an increase in $ xH_v^u(x) $ at small values of $ x $ (note that the peak of valence GPDs, e.g., occurs at $ x=0.4 $ for $ t=-3 $ GeV$ ^2 $), and a minor decrease at $ x\simeq 0.1-0.2 $. The results also show that without considering the Mainz data, the impact of GMp12 on $ xH_v^u(x) $ at small $ x $ region becomes even stronger. An interesting finding is the significant reduces in uncertainties of $ xH_v^u(x) $ at $ x\gtrsim 0.02 $ especially as the absolute value of $ -t $ increases. Note that the large $ x $ region is correlated to the large values of $ t $. This clearly indicates the crucial role of the GMp12 data in the future analysis of GPDs. 

\begin{figure}[!htb]
    \centering
\includegraphics[scale=1.3]{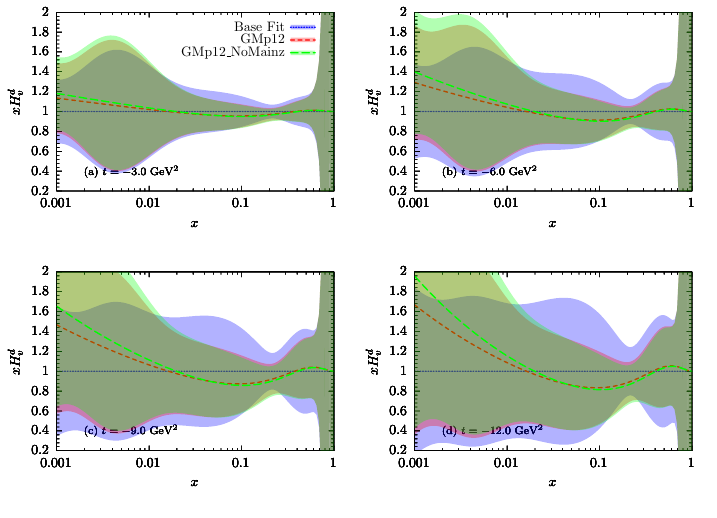}   
    \caption{Same as Fig.~\ref{fig:Huv}, but for valence unpolarized GPD  $ xH_v^d(x) $.}
\label{fig:Hdv}
\end{figure}

Figure~\ref{fig:Hdv} shows the same results as Fig.~\ref{fig:Huv}, but for the valence unpolarized GPD $ xH_v^d(x) $. The results obtained show that the impact of GMp12 data on the down quark distribution is almost similar to the case of up quark considering the trends, though it is somewhat stronger in magnitude.  
Although the GMp12 data are led to a decrease in uncertainties of $ xH_v^d(x) $ at $ x\gtrsim 0.02 $, the severity of this decrease is less compared to the case of $ xH_v^u(x) $. This was predictable
considering the fact that GMp12 data are pertained to the electron scattering off the proton. While for a remarkable reduce in uncertainties of the down quark distribution we need precise measurements of the neutron (or deuteron) cross-section. 

\begin{figure}[!htb]
    \centering
\includegraphics[scale=1.3]{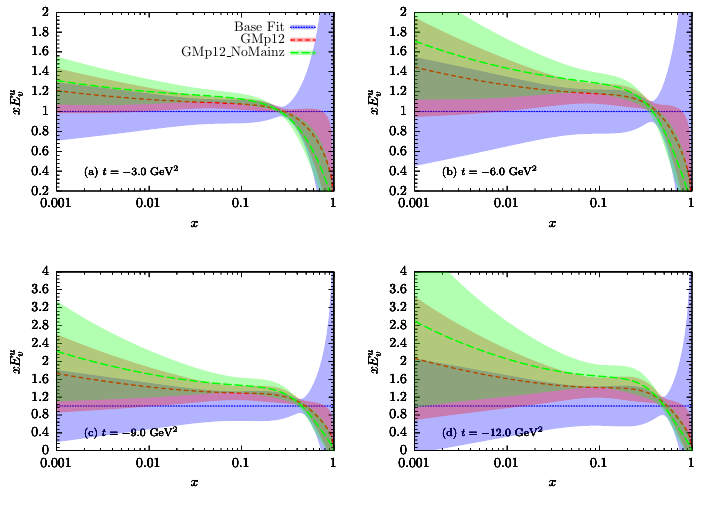}   
    \caption{Same as Fig.~\ref{fig:Huv}, but for valence unpolarized GPD $ xE_v^u(x) $.}
\label{fig:Euv}
\end{figure}

The results obtained for GPD $ xE_v^u(x) $ are shown in Fig.~\ref{fig:Euv} and compared again at $ t=-3,-6,-9,-12 $ GeV$ ^2 $. Note that, as mentioned before, there is no equivalent of PDFs for the forward limits of GPDs $ E_v^q $. So, in this case, both the forward limits  $ e_v^q(x) $ and profile functions $ g_v^q(x) $ are obtained from the fit to data. In this way, $ xE_v^u(x) $ distributions (and also $ xE_v^d(x) $) obtained from three analyses performed in this study are different even at $ t=0 $. However, since the differences increase as the absolute value of $ -t $ increases, we have not shown here the results corresponding to $ t=0 $. As can be seen, $ xE_v^u(x) $ has been affected more after the inclusion of the GMp12 data in the analysis compared with $ xH_v^u(x) $ in Fig.~\ref{fig:Huv}. It is logical because  according to Eq.~(\ref{Eq5}) as the value of $ Q^2 $ ($ -t $) increases, the contribution of $ F_2 $ which contain GPDs $ E_v^q $ becomes grater. Note also that compared with GPD $ xH_v^u(x) $, $ xE_v^u(x) $ has been impacted almost at all values  of $ x $. An egregious difference is that $ xE_v^u(x) $ undergoes a drastic suppression at large $ x $ region. 
These findings clearly indicate that considering the GMp12 data in the global analysis of GPDs at zero skewness can significantly affect both the shape and uncertainty of GPD $ xE_v^u(x) $.

\begin{figure}[!htb]
    \centering
\includegraphics[scale=1.3]{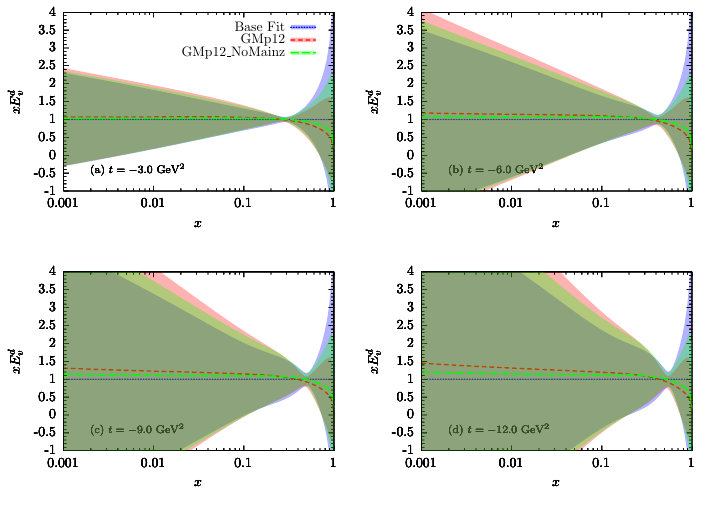}
    \caption{Same as Fig.~\ref{fig:Huv} but for valence unpolarized GPD $ xE_v^d(x) $. }
\label{fig:Edv}
\end{figure}

Figure~\ref{fig:Edv} shows the same results as Fig.~\ref{fig:Huv} but for GPD $ xE_v^d(x) $. 
According to the results obtained, one can conclude that the GMp12 data do not provide any constrain on GPD
$ xE_v^d(x) $ at small and medium values of $ x $. However, some changes in $ xE_v^d(x) $ are seen at large $ x $ region both in shape and uncertainty. It should be noted again that GMp12 data have been measured from the scattering off the proton and it is sensible that they can not significantly affect the down quark distribution especially considering the smaller contributions of GPDs $ E_v^q $ rather that $ H_v^q $ in the total cross-section.  
%

%

\section{Nucleon to $ \Delta $ electromagnetic transition}\label{sec:four} 

Alongside the usual FFs of the nucleon, their transition FFs to another particle can also provide valuable information about the internal structure and dynamics of the nucleons~\cite{Stoler:1993yk,Burkert:2004sk,Aznauryan:2011qj,Aznauryan:2012ba,Hagelstein:2017cbl,Proceedings:2020fyd,Burkert:2022ioj,Ramalho:2023hqd}. Actually, TFFs are crucial in understanding processes like electromagnetic and weak interactions in particle physics. By investigating how TFFs are changed during particle interactions or decays, one can gain insights into fundamental aspects of particle behavior, such as their charge distribution, momentum transfer, and the strength of their interactions with other particles. This knowledge is essential for developing and refining theoretical models, as well as for interpreting experimental data. Overall, there are several types of TFFs~\cite{Ramalho:2023hqd,Aung:2024qmf,Chen:2023zhh,Ozdem:2022zig,Lu:2019bjs} including  the electromagnetic TFFs (EM TFFs), weak FFs, gravitational TFFs, and axial TFFs. Each type of form factor provides information about different aspects of particle interactions and properties. Among them, EM TFFs~\cite{Burkert:2022ioj,Ramalho:2023hqd,Aung:2024qmf} provides information about how the particle interacts electromagnetically with other particles or fields. They quantify how the charge is distributed within the particle and how it responds to electromagnetic interactions. In the experimental point of view, EM TFFs are measured by studying the scattering or decay of particles involving electromagnetic interactions, such as photon emission or absorption.

Measurements of the nucleon-to-delta ($ N\rightarrow \Delta $) electromagnetic TFFs
are typically performed by utilizing electron scattering off a nucleon target ($ \Delta $ can also be produced in scattering pions off a nucleon). By measuring the differential cross-sections and polarization observables of the scattered electrons, one can extract information about the TFFs. Experimental facilities such as Jefferson Lab in the US, MAMI in Germany, and others around the world have conducted experiments to precisely measure these TFFs. It is well established now that the $ \Delta(1232) $ resonance is the first excited state of the nucleon and has isospin $ 3/2 $, and therefore can be existed in four different charge states: $ \Delta^{++} $, $ \Delta^{+} $, $ \Delta^{0} $, and $ \Delta^{-} $. These states have approximately the same mass and width. Note that the spin of the $ \Delta(1232) $ is also $ 3/2 $.

The experiments in the region of the $ \Delta(1232) $ resonance are often used to measure the electric and scalar quadrupole ratios, $ R_{EM} $ and $ R_{SM} $, which are defined as~\cite{Burkert:2022ioj}
\begin{equation}
\label{Eq13}
R_{EM}\equiv \frac{Im(E_{1+})}{Im(M_{1+})}\,, ~~~~~~~~ R_{SM}\equiv \frac{Im(S_{1+})}{Im(M_{1+})}\,,
\end{equation}
where $ E_{1+} $, $ M_{1+} $, and $ S_{1+} $ are the electromagnetic transition multipoles at the mass of the $ \Delta(1232) $ resonance. The multipole ratios can also be written in terms
of the Jones-Scadron FFs as~\cite{Hagelstein:2017cbl}
\begin{equation}
\label{Eq14}
R_{EM}(Q^2)= -\frac{G^*_E(Q^2)}{G^*_M(Q^2)}\,, ~~~~~~~~ R_{SM}(Q^2)= -\frac{Q_+ Q_-}{4M^2_{\Delta}} \frac{G^*_C(Q^2)}{G^*_M(Q^2)}\,,
\end{equation}
where $ G^*_E $, $ G^*_M $, and $ G^*_C $  are the nucleon-to-delta TFFs. These TFFs can be connected to the electromagnetic nucleon properties via large number of color ($ N_c $) relations and modeled at finite momentum transfer as~\cite{Pascalutsa:2007wz}
\begin{align}
{\label{Eq15}}
G^*_M(Q^2)=&\frac{1}{\sqrt{2}} \left[ F_{2}^p (Q^2)-F_{2}^n (Q^2) \right]\,,\nonumber\\
G^*_E(Q^2)=& \left( \frac{M}{M_{\Delta}} \right) ^{3/2} \frac{\Delta M_+}{2\sqrt{2} Q^2} G_{E}^n(Q^2)\,,\nonumber\\
G^*_C(Q^2)=&\frac{4M_{\Delta}^2}{\Delta M_+} G^*_E(Q^2)\,.
\end{align}  
In above equations, $ \Delta M_{+}= M_{\Delta}^2-m^2 $, where $ M_{\Delta} $ and $ m $ denote the mass of $ \Delta(1232) $ and nucleon respectively. Considering Eqs.~(\ref{Eq5}),~(\ref{Eq6}), and~(\ref{Eq15}) together, one can conclude that the nucleon-to-delta TFFs $ G^*_E $, $ G^*_M $, and $ G^*_C $ are also related to the valence unpolarized GPDs $ H_v^q $ and $ E_v^q $. Hence, by measuring these TFFs, or equivalently the electric and scalar quadrupole ratios, $ R_{EM} $ and $ R_{SM} $, one can get new and important information on GPDs at zero skewness. Fortunately, there have been various measurements of $ R_{EM} $ and $ R_{SM} $ (note that TFFs can also be extracted in such measurements) up to now. So, it is interesting to see that how GPDs extracted in the previous section through the analysis of the electromagnetic FFs describe the available data relating to the nucleon-to-delta transition.

\begin{figure}[!htb]
    \centering
\includegraphics[scale=0.6]{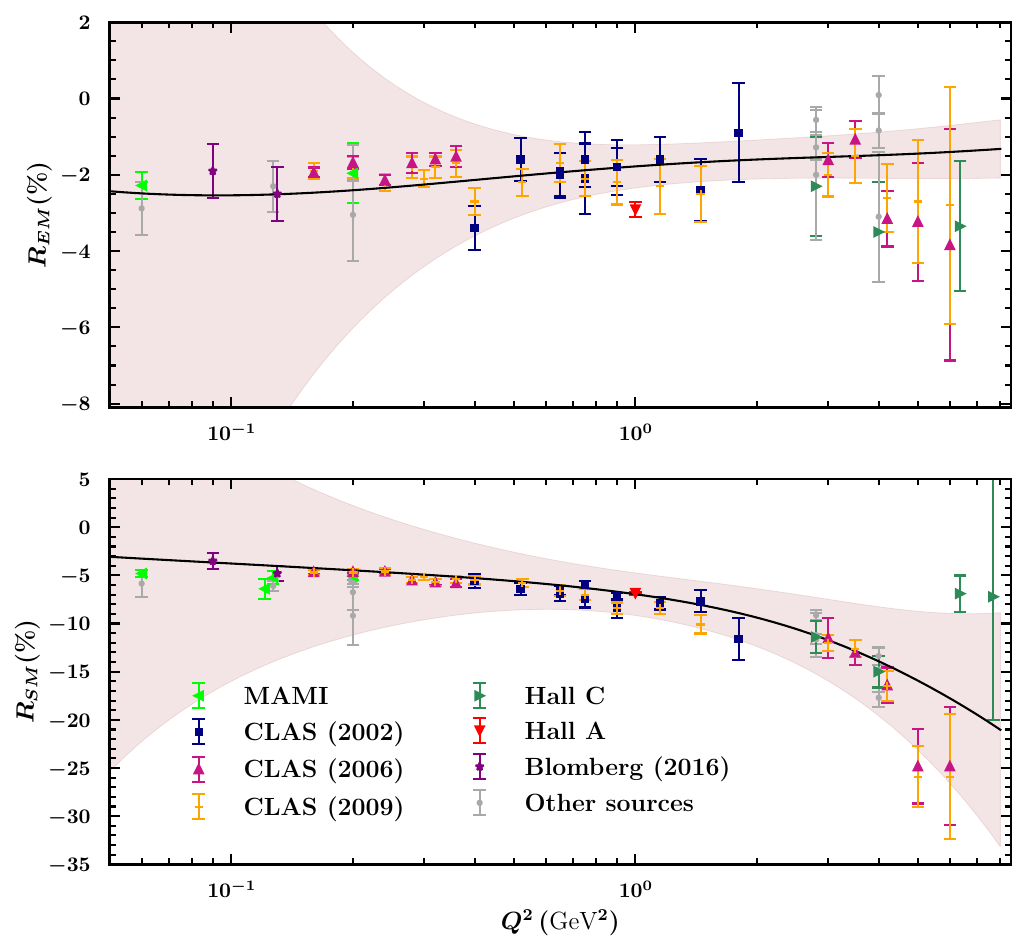}    
    \caption{A comparison between our results for $ R_{EM} $ and $ R_{SM} $ (expressed in percent) obtained using the set of GPDs named GMp12 and the corresponding experimental measurements from MAMI~\cite{Stave:2006ea,Sparveris:2006uk,Sparveris:2013ena,Pospischil:2000ad}, CLAS (2002)~\cite{CLAS:2001cbm}, CLAS (2006)~\cite{CLAS:2006ezq,Julia-Diaz:2006ios}, CLAS (2009)~\cite{CLAS:2009ces}, Hall C~\cite{Frolov:1998pw,Villano:2009sn}, Hall A~\cite{JeffersonLabHallA:2005wfu}, and Blomberg (2016)~\cite{Blomberg:2015zma}. The other experimental data denoted as ``Other sources" have been gathered from Ref.~\cite{A1:2008ocu,Elsner:2005cz,OOPS:2004kai,Tiator:2000iy}.}
\label{fig:RESM}
\end{figure}

Figure~\ref{fig:RESM} shows a comparison between the experimental measurements of $ R_{EM} $ (top panel) and $ R_{SM} $ (bottom panel) from various experiments~\cite{Stave:2006ea,Sparveris:2006uk,Sparveris:2013ena,Pospischil:2000ad,CLAS:2001cbm,CLAS:2006ezq,Julia-Diaz:2006ios,CLAS:2009ces,Frolov:1998pw,Villano:2009sn,JeffersonLabHallA:2005wfu,Blomberg:2015zma,A1:2008ocu,Elsner:2005cz,OOPS:2004kai,Tiator:2000iy} and the corresponding theoretical prediction calculated using the GPDs set named GMp12 obtained in the previous section. Note that we have presented the results in percent as it is common. In this plot, we have separated the data of CLAS (2002)~\cite{CLAS:2001cbm}, CLAS (2006)~\cite{CLAS:2006ezq,Julia-Diaz:2006ios}, CLAS (2009)~\cite{CLAS:2009ces}, Hall C~\cite{Frolov:1998pw,Villano:2009sn} and Hall A~\cite{JeffersonLabHallA:2005wfu} from JLab as well as MAMI~\cite{Stave:2006ea,Sparveris:2006uk,Sparveris:2013ena,Pospischil:2000ad} and Blomberg (2016)~\cite{Blomberg:2015zma}. The other experimental data denoted as ``Other sources" have been gathered from Refs.~\cite{A1:2008ocu,Elsner:2005cz,OOPS:2004kai,Tiator:2000iy}. Note also that there are no remarkable differences between the predictions of three sets of GPDs obtained in the previous section. So, we decided to include only the results of the set named GMp12 that has been obtained by considering all data introduced in Sec.~\ref{sec:three-two}. As can be seen from Fig.~\ref{fig:RESM},  our results are in a very good consistency with the experimental data. Note again that these data have not included in our analyses of GPDs. The good agreement between the theoretical predictions and experimental measurements beautifully proves the universality property of GPDs. Another important point can be deduced from Fig.~\ref{fig:RESM} is that the inclusion of the $ R_{EM} $ and $ R_{SM} $ data in the analysis of GPDs at zero skewness can significantly impacted the extracted GPDs especially their uncertainties at smaller values of $ Q^2 $. Consequently, we strongly suggest renewing the comprehensive analysis performed in Ref.~\cite{Hashamipour:2022noy} by considering also the $ R_{EM} $ and $ R_{SM} $ data in the analysis. 

\begin{figure}[!htb]
    \centering
\includegraphics[scale=0.8]{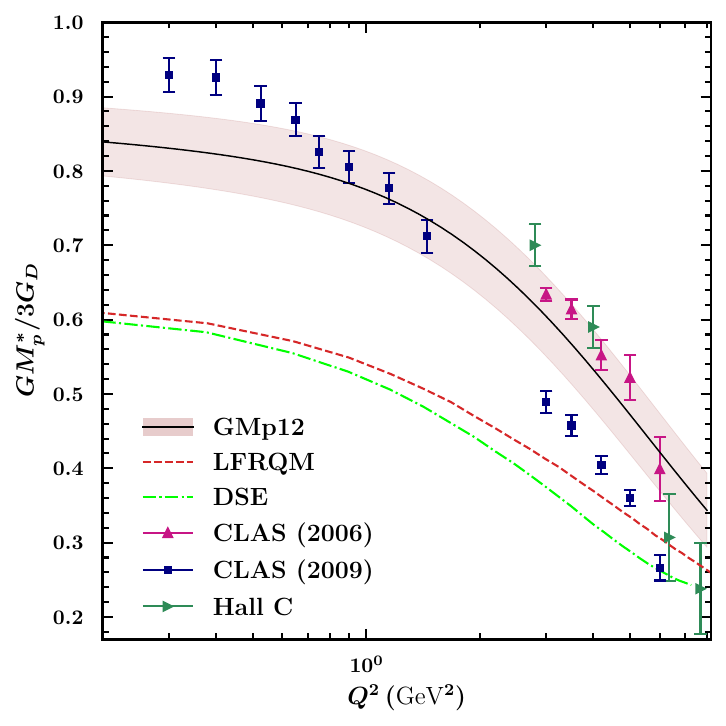}    
    \caption{A comparison between our result for $ G^*_M/3G_D $ obtained using the set of GPDs named GMp12 and the corresponding experimental measurements from CLAS (2006)~\cite{CLAS:2006ezq,Julia-Diaz:2006ios}, CLAS (2009)~\cite{CLAS:2009ces}, and Hall C~\cite{Frolov:1998pw,Villano:2009sn} as well as the pure theoretical calculations from the LFRQM~\cite{Aznauryan:2012ec,Aznauryan:2016wwm} and the Dyson-Schwinger equation~\cite{Segovia:2014aza}.}
\label{fig:GMst}
\end{figure}

It is also of interest to compare the results for the magnetic transition form factor $ G^*_M $ which is a very sensitive quantity to the valence unpolarized GPDs $ E_v^q $. Note that our understanding of GPDs $ E_v^q $ have always been less robust compared with that of GPDs $ H_v^q $ because of fewer constraints on $ E_v^q $ from the experimental measurements~\cite{Hashamipour:2022noy,Hashamipour:2021kes}. In Fig.~\ref{fig:GMst}, we have presented a comparison between the experimental measurements of $ G^*_M/3G_D $ from CLAS (2006)~\cite{CLAS:2006ezq}, CLAS (2009)~\cite{CLAS:2009ces}, and Hall C~\cite{Frolov:1998pw,Villano:2009sn} and the corresponding theoretical prediction obtained using GPDs set GMp12 as well as the Light-Front Relativistic Quark Model (LFRQM)~\cite{Aznauryan:2012ec,Aznauryan:2016wwm} and the Dyson-Schwinger Equation (DSE)~\cite{Segovia:2014aza}. It should be noted that the data points from CLAS (2006) and Hall C have been presented in their related references, but for CLAS (2009) we have calculated them according to the information presented in~\cite{CLAS:2009ces}. As can be seen from Fig.~\ref{fig:GMst}, our result is again in a sensible consistency with the experimental data especially compared with the results of LFRQM and DSE. The results obtained clearly confirm this statement that the inclusion of the measurements of the TFFs in the analysis of GPDs at zero skewness can play a crucial role to constrain them and reduce their uncertainties especially at small $ -t $ region.
%

%

\section{Summary and conclusion}\label{sec:five} 
 
In this study, we showed that the JLab GMp12 data~\cite{Christy:2021snt} of the reduced cross-section of the elastic electron-proton scattering play a crucial role in disentangling the tension observed in Ref.~\cite{Hashamipour:2022noy} between the world data of $ G_M^p $ from AMT07 analysis~\cite{Arrington:2007ux} and the corresponding ones from Mainz~\cite{A1:2013fsc} as well as that observed between the measurements of the WACS cross-section~\cite{Danagoulian:2007gs} and $ G_M^p $ data. As a result, we found that GMp12 data are more consistent with AMT07 $ G_M^p $ world data rather than Mainz data. They also do not recommend the exclusion of the $ G_M^p $ data from the analysis, while the WACS data do this. This may suggest to revising the theoretical calculations relating to the WACS cross-section or improving the phenomenological framework.

As the next step, we investigated the impact of GMp12 data on valence unpolarized GPDs $ H_v^q $ and $ E_v^q $ at zero skewness by performing two separate analyses, one with and the other without considering the GMp12 data in the analysis, and comparing their results with each other. As expected, we found that GMp12 data have more impact on GPDs as the absolute value of $ -t $ increases. They lead to an increase in $ xH_v^u(x) $ and $ xH_v^d(x) $ at small values of $ x $ as well as a minor decrease at medium $ x $ region. For the case of GPD $ xE_v^u(x) $, we found that it has been affected more compared to $ xH_v^u(x) $ so that it undergoes a drastic suppression at large $ x $ region too. However, we indicated that the GMp12 data do not provide remarkable constrains on GPD $ xE_v^d(x) $ except at large $ x $ region. Overall, the results show that by excluding the Mainz data from the analysis, the impact of GMp12 on GPDs at small $ x $ region becomes even stronger. 
A very important finding of the present study is the significant reduces in uncertainties of GPDs at $ x\gtrsim 0.02 $, especially as the absolute value of $ -t $ increases, after the inclusion of the GMp12 data in the analysis. These findings clearly indicate the crucial role of the GMp12 data in the future analysis  of GPDs at zero skewness. We emphasis that considering these data can significantly affect both the shape and uncertainty of GPDs.

By calculating the electric and scalar quadrupole ratios, $ R_{EM} $ and $ R_{SM} $, of the the $ \Delta(1232) $ resonance and the nucleon-to-delta magnetic transition form factor $ G^*_M/3G_D $ theoretically using the extracted GPDs and comparing the results with corresponding experimental measurements, we showed that there is an excellent consistency between our results and experiments 
that beautifully proves the universality property of GPDs. Our predictions have also much better description of data compared with the pure theoretical prediction obtained from LFRQM and DSE.
We indicated that the inclusion of these data in the analysis of GPDs at zero skewness can significantly impacted the extracted GPDs especially their uncertainties at smaller values of $ Q^2 $. Consequently, we strongly suggest renewing the comprehensive analysis performed in Ref.~\cite{Hashamipour:2022noy} by considering also the experimental data relating to the $ N\rightarrow \Delta $ transition. This aspect will be further explored in our future research.

%
\section*{ACKNOWLEDGMENTS}
M.~Goharipour and H.~Hashamipour are thankful to the School of Particles and Accelerators, Institute for Research in Fundamental Sciences (IPM), for financial support provided for this research. M.~Goharipour thanks the Theoretical Physics Department, CERN, for providing financial support for a short-term visit during which a crucial part of the present project was done.

%
\section*{Note Added}
 We are pleased to provide  the GPDs extracted in the present study with their uncertainties in any desired
values of $ x $ and $ -t $ upon request.

%


\end{document}